\documentclass{aa}
\usepackage{graphics}
 
\begin{document}      

   \title{Atomic gas far away from the Virgo cluster core galaxy NGC~4388}

   \subtitle{A possible link to isolated star formation in the Virgo cluster?}

   \author{B.~Vollmer, W. Huchtmeier}

   \offprints{B.~Vollmer, e-mail: bvollmer@mpifr-bonn.mpg.de}

   \institute{Max-Planck-Institut f\"ur Radioastronomie, Auf dem H\"ugel 69,
              D-53121 Bonn, Germany}

   \date{Received / Accepted}

   \authorrunning{Vollmer \& Huchtmeier}
   \titlerunning{Atomic gas far away from NGC~4388}

\abstract{
We have discovered 6\,10$^{7}$~M$_{\odot}$ of atomic gas at a projected distance greater than
4$'$ (20~kpc) from the highly inclined Virgo spiral galaxy NGC~4388. This gas is most
probably connected to the very extended H$\alpha$ plume detected by Yoshida et al. (2002).
Its mass makes a nuclear outflow and its radial velocity a minor merger as
the origin of the atomic and ionized gas very unlikely. 
A numerical ram pressure simulation can account for the observed H{\sc i} spectrum and the 
morphology of the H$\alpha$ plume. An additional outflow mechanism is still 
needed to reproduce the velocity field of the inner H$\alpha$ plume.
The extraplanar compact H{\sc ii} region recently found by Gerhard et al. (2002) can be 
explained as a stripped gas cloud that collapsed and decoupled from the ram pressure wind
due to its increased surface density. The star-forming cloud is now falling back onto the galaxy.
}

\maketitle

\section{Introduction \label{sec:intro}}

The Virgo cluster spiral galaxy NGC~4388 is located at a projected distance of 
1.3$^{\rm o}$ from the Virgo cluster center (M87). Its high radial velocity
of $\sim$1400~km\,s$^{-1}$ with respect to the cluster mean and its estimated
line-of-sight distance (Tully-Fisher method; Yasuda et al. 1997) 
to the cluster center place its three dimensional location very close to M87. 
In addition, NGC~4388 represents one of the nearest Seyfert~2
galaxies and the first AGN to be found in the Virgo cluster (Phillips \&
Malin 1982). Veilleux et al. (1999) made observations with a Fabry-Perot 
Interferometer in the H$\alpha$ and O[{\sc iii}] $\lambda$5007 lines.
They found a large, high-ionization plume extending northeastwards from the 
nucleus up to a projected distance of 4~kpc\footnote{We adopt a distance of
17~Mpc to the Virgo cluster.} above the plane of the galaxy.
The plume is blueshifted with respect to the galaxy's systemic velocity.
Yoshida et al. (2002) discovered a very large H$\alpha$ plume that extends
up to $\sim$35~kpc northeastwards. This region contains $\sim 10^{5}$~M$_{\odot}$
of ionized gas.
They argued that the dominating source of ionization is the radiation coming
from the nucleus. Veilleux et al. (1999) discussed 4 different origins for the
4~kpc plume: (i) a minor merger, (ii) a galactic wind, (iii) nuclear outflow,
and (iv) ram pressure stripping. Based on their data they favoured a combination 
of (iii) and (iv). Yoshida et al. (2002) on the other hand favoured scenario (i) and
(iv). A surprisingly new result is that Gerhard et al. (2002) found an
isolated compact H{\sc ii} region at a projected distance of 17~kpc north and
4.4~kpc west of the galaxy center. The age of the dozen O stars embedded
in this region ($\sim$3~Myr) implies that they
have formed far outside the main star-forming regions of the galaxy
and perhaps even within the hot intracluster medium. Gerhard et al. (2002)
argued that a radial velocity of 150~km\,s$^{-1}$ excludes that the involved gas has been 
pushed to this large  distance by ram pressure stripping. In addition to the
studied isolated compact H{\sc ii} region they found 17 H{\sc ii} region candidates 
in the data of Yoshida et al. (2002).
In this article we report the detection of atomic gas at a projected distance
of more than 20~kpc northeastwards from the center of NGC~4388.
The observations and their results are described in Sect.~\ref{sec:observations} and
Sect.~\ref{sec:results}. The dynamical model is presented in Sect.~\ref{sec:model}.
We compare a model snapshot with our observations and discuss possible implications
in Sect.~\ref{sec:discussion}. The conclusions are given in Sect.~\ref{sec:conclusions}.

\section{Observations \label{sec:observations}}

On March, 14--21 2002 we performed 21-cm line observations at 5 different positions at a frequency 
corresponding to the systemic velocity of NGC~4388 with a bandwidth of 12.5~MHz. 
The two channel receiver had a system noise of $\sim$30~K. The 1024 channel autocorrelator
was split into four parts with 256 channels, yielding a
channel separation of $\sim$10~km\,s$^{-1}$. We further binned the channels to obtain
a final channel separation of $\sim$20~km\,s$^{-1}$. One central position and four 
positions at a distance of one beamsize (9.3$'$) to the NW, SW, SE, and
NE from the galaxy center were observed in on--off mode
(5~min on source, 5~min off source). The integration times are shown in Table~\ref{tab:table}.
\begin{table}
      \caption{Integration times and rms.}
         \label{tab:table}
      \[
         \begin{array}{lccccc}
           \hline
           \noalign{\smallskip}
           {\rm position} & {\rm Center} & {\rm Northwest} & {\rm Southwest} & {\rm Southeast} & {\rm Northeast} \\
	   \noalign{\smallskip}
	   \hline
	   \noalign{\smallskip}
	   $$\Delta t$$\ {\rm (min)} & 190 & 80 & 120 & 120 & 350 \\ 
           \noalign{\smallskip}
	   \hline
	   \noalign{\smallskip}
	   {\rm rms\ (mJy)} & 1.2 & 1.7 & 1.1 & 1.2 & 0.9 \\
	   \noalign{\smallskip}
        \hline
        \end{array}
      \]
\end{table}
Care was taken to avoid other Virgo galaxies with velocities within our bandwidth
in all 5 on and off source positions. We used 3C286 for pointing and flux
calibration. The observation times ranged between 80~min and 350~min per position.
The resulting noise (Table~\ref{tab:table}) is partly determined by small amplitude 
interferences, but is close to the theoretical noise of 2~mJy per hour of integration.

\section{Results \label{sec:results}}

\begin{figure}
        \resizebox{\hsize}{!}{\includegraphics{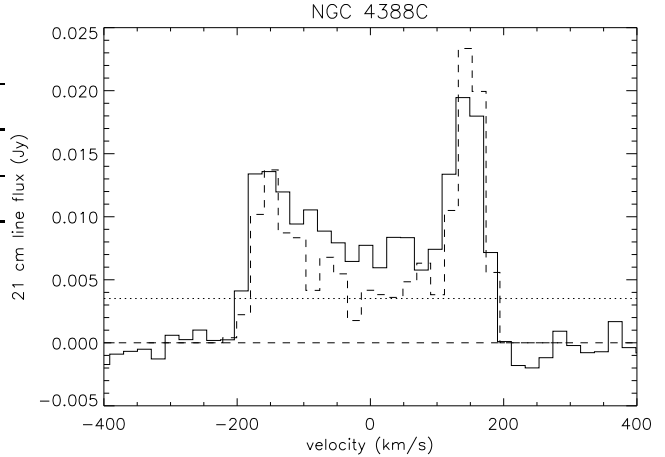}}
        \caption{Solid line: Effelsberg 100-m spectrum of the central position.
	Dashed line: spectrum of the VLA data. Dotted line: 3$\sigma$ noise level of the
	100-m spectrum. Heliocentric velocities are given relative to the systemic velocity
	of NGC~4388 ($v_{\rm sys}$=2524~km\,s$^{-1}$).
        } \label{fig:n4388c}
\end{figure}

The Effelsberg 100-m spectrum of the central position can be seen in Fig.~\ref{fig:n4388c}
(solid line). It shows a double-horn structure with a $\sim$25\% higher peak emission
at the receding side. The total flux is $S_{\rm HI}$=3.9~Jy\,km\,s$^{-1}$, which corresponds to 
a total H{\sc i} mass of $M_{\rm HI}=2.7\,10^{8}$~M$_{\odot}$. 

In order to compare our off-center positions to interferometric data where the
galaxy is spatially resolved, we use VLA 21~cm C array data (Cayatte et al. 1990). 
These data have a spatial resolution of $23'' \times 17''$ and a channel separation
of 20~km\,s$^{-1}$. We clipped the data cube at a level of 4~mJy/beam and cut out
the inner $4' \times 3'$ around the galaxy center. The resulting
spectrum for the central position is represented by the dashed line in Fig.~\ref{fig:n4388c}.
The total flux and peak flux density of Cayatte et al. (1990) are $\sim$20\% higher than our values, 
which is within the calibration accuracy. Both spectra agree reasonably well.

\begin{figure*}
        \resizebox{\hsize}{!}{\includegraphics{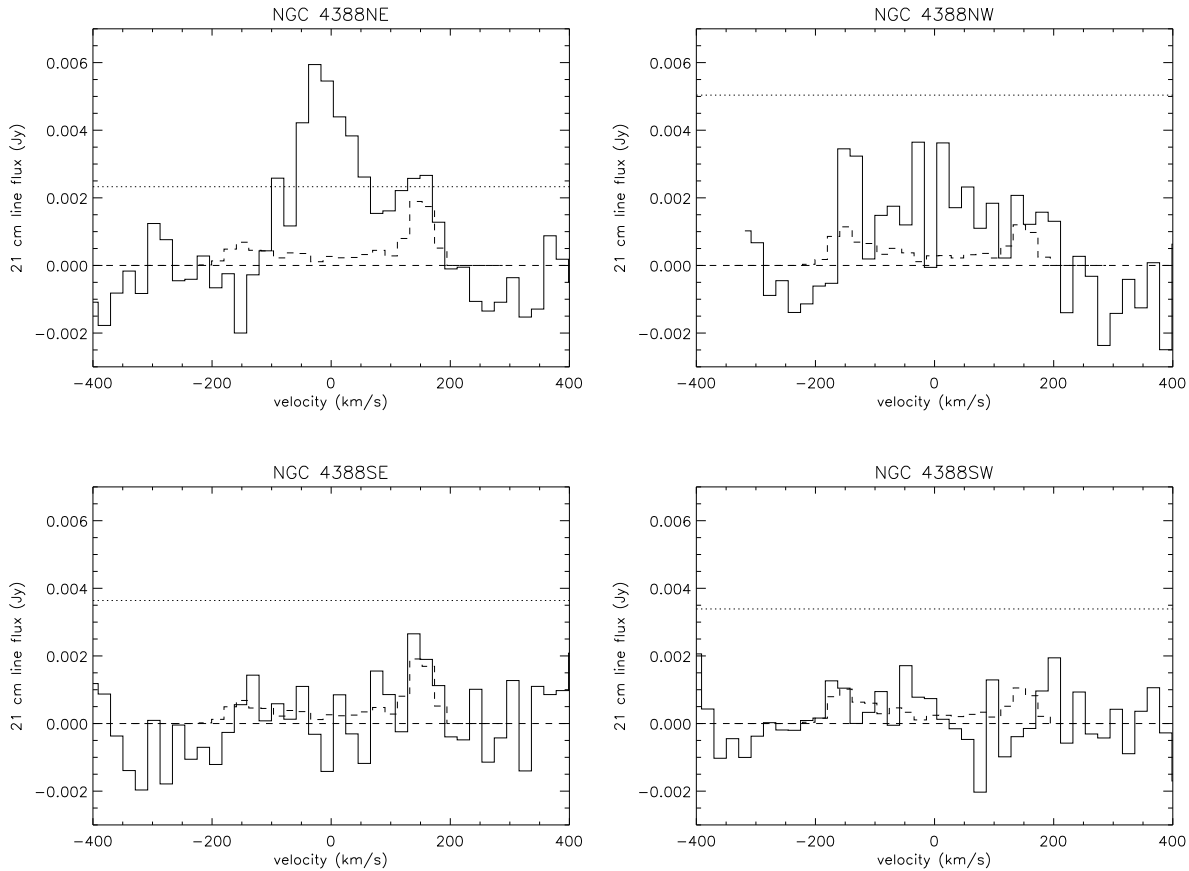}}
        \caption{Solid lines: Effelsberg 100-m spectra of the four off-center positions.
	Their locations with respect to the galaxy center are marked on top of each panel.
	Dashed line: synthesized VLA spectra which only show H{\sc i} disk emission. 
	Dotted line: 3$\sigma$ noise levels of the
	100-m spectra. Velocities are given relative to the systemic velocity of NGC~4388.
        } \label{fig:n4388_spectra}
\end{figure*}

For the 4 off-center positions we have synthesized spectra from the VLA data using the position of
the pointing and the beamsize of the Effelsberg 100-m telescope at 21~cm (9.3$'$).
We use the spatial information of the VLA data to construct the synthesized spectra.
Fig.~\ref{fig:n4388_spectra} shows the 100-m spectra of these positions (solid lines)
together with the synthesized VLA spectra. The VLA spectra represent H{\sc i} gas
located in the disk (see Cayatte et al. 1990). If there is an emission excess
in the off-center 100-m spectra this must be due to gas located outside the disk.
The rms noise of the north-eastern (NE) position is certainly overestimated, because
of a relatively strong standing wave that had to be subtracted.
We made 3 independent observations of this position in 3 different nights.
In all 3 sub-spectra the signal is detected. One of them shows a flat baseline
without a standing wave. 
The peak emission of none of the synthesized VLA spectra exceeds the 100-m 3$\sigma$ noise level,
except the NE position where we have overestimated the rms.

Only the 100-m spectrum of the NE position shows a detection. The main peak is around
zero radial velocity with respect to the galaxy. It has a width of $\sim$80~km\,s$^{-1}$.
In addition, we detect a second smaller maximum at 160~km\,s$^{-1}$, which has the same
position and width as the corresponding emission of the eastern, receding side of NGC~4388.
However, it contains $\sim$30\% more flux than the corresponding maximum of the VLA data.
The total flux of the emission is $S_{\rm HI}$=0.9~Jy\,km\,s$^{-1}$, which corresponds 
to an atomic gas mass of $M_{\rm HI}=6\,10^{7}$~M$_{\odot}$.

\section{The model \label{sec:model}}

Since the model is described in detail in Vollmer et al. (2001), we 
summarize only its main features.
The N-body code consists of two components, a non-collisional component
that simulates the stellar bulge/disk and the dark halo, and a
collisional component that simulates the ISM.

The 20\,000 particles of the collisional component represent gas cloud complexes which are 
evolving in the gravitational potential of the galaxy.

The total gas mass is $M_{\rm gas}^{\rm tot}=3.8\,10^{9}$~M$_{\odot}$,
which corresponds to the total neutral gas mass before stripping
assuming an H{\sc i} deficiency of 1.1.
To each particle a radius is attributed depending on its mass. 
During the disk evolution the particles can have inelastic collisions, 
the outcome of which (coalescence, mass exchange, or fragmentation) 
is simplified following Wiegel (1994). 
This results in an effective gas viscosity in the disk. 

As the galaxy moves through the ICM, its clouds are accelerated by
ram pressure. Within the galaxy's inertial system its clouds
are exposed to a wind coming from the opposite direction of the galaxy's 
motion through the ICM. 
The temporal ram pressure profile has the form of a Lorentzian,
which is realistic for galaxies on highly eccentric orbits within the
Virgo cluster (Vollmer et al. 2001).
The effect of ram pressure on the clouds is simulated by an additional
force on the clouds in the wind direction. Only clouds which
are not protected by other clouds against the wind are affected.

The non--collisional component consists of 49\,125 particles, which simulate
the galactic halo, bulge, and disk.
The characteristics of the different galactic components are shown in
Table~\ref{tab:param}.
\begin{table}
      \caption{Total mass, number of particles $N$, particle mass $M$, and smoothing
        length $l$ for the different galactic components.}
         \label{tab:param}
      \[
         \begin{array}{lllll}
           \hline
           \noalign{\smallskip}
           {\rm component} & M_{\rm tot}\ ({\rm M}$$_{\odot}$$)& N & M\ ({\rm M}$$_{\odot}$$) & l\ ({\rm pc}) \\
           \hline
           {\rm halo} & 2.4\,10$$^{11}$$ & 16384 & $$1.4\,10^{7}$$ & 1200 \\
           {\rm bulge} & 8.2\,10$$^{9}$$ & 16384 & $$5.0\,10^{5}$$ & 180 \\
           {\rm disk} & 4.1\,10$$^{10}$$ & 16384 & $$2.5\,10^{6}$$ & 240 \\
           \noalign{\smallskip}
        \hline
        \end{array}
      \]
\end{table}
The resulting rotation velocity is $\sim$200~km\,s$^{-1}$ and the rotation curve
is flat. 

The particle trajectories are integrated using an adaptive timestep for
each particle. This method is described in Springel et al. (2001).
The following criterion for an individual timestep is applied:
\begin{equation}
\Delta t_{\rm i} = \frac{20~{\rm km\,s}^{-1}}{a_{\rm i}}\ ,
\end{equation}
where $a_{i}$ is the acceleration of the particle i.
The minimum of all $t_{\rm i}$ defines the global timestep used 
for the Burlisch--Stoer integrator that integrates the collisional
component.

The galaxy is on an eccentric orbit within the cluster. The temporal
ram pressure profile can be described by:
\begin{equation}
p_{\rm ram}=\frac{p_{\rm max}}{t^{2}+t_{\rm HW}^{2}}\ ,
\end{equation}
where $t_{\rm HW}$ is the width of the profile (Vollmer et al. 2001). 
We set $p_{\rm max}$=5000~cm$^{-3}$(km/s)$^{2}$ and $t_{\rm HW}$=50~Myr.
The efficiency of ram pressure also depends on the inclination angle $i$
between the galactic disk and the orbital plane (Vollmer et al. 2001).
We set $i$=45$^{\rm o}$.

\begin{figure*}
        \resizebox{\hsize}{!}{\includegraphics{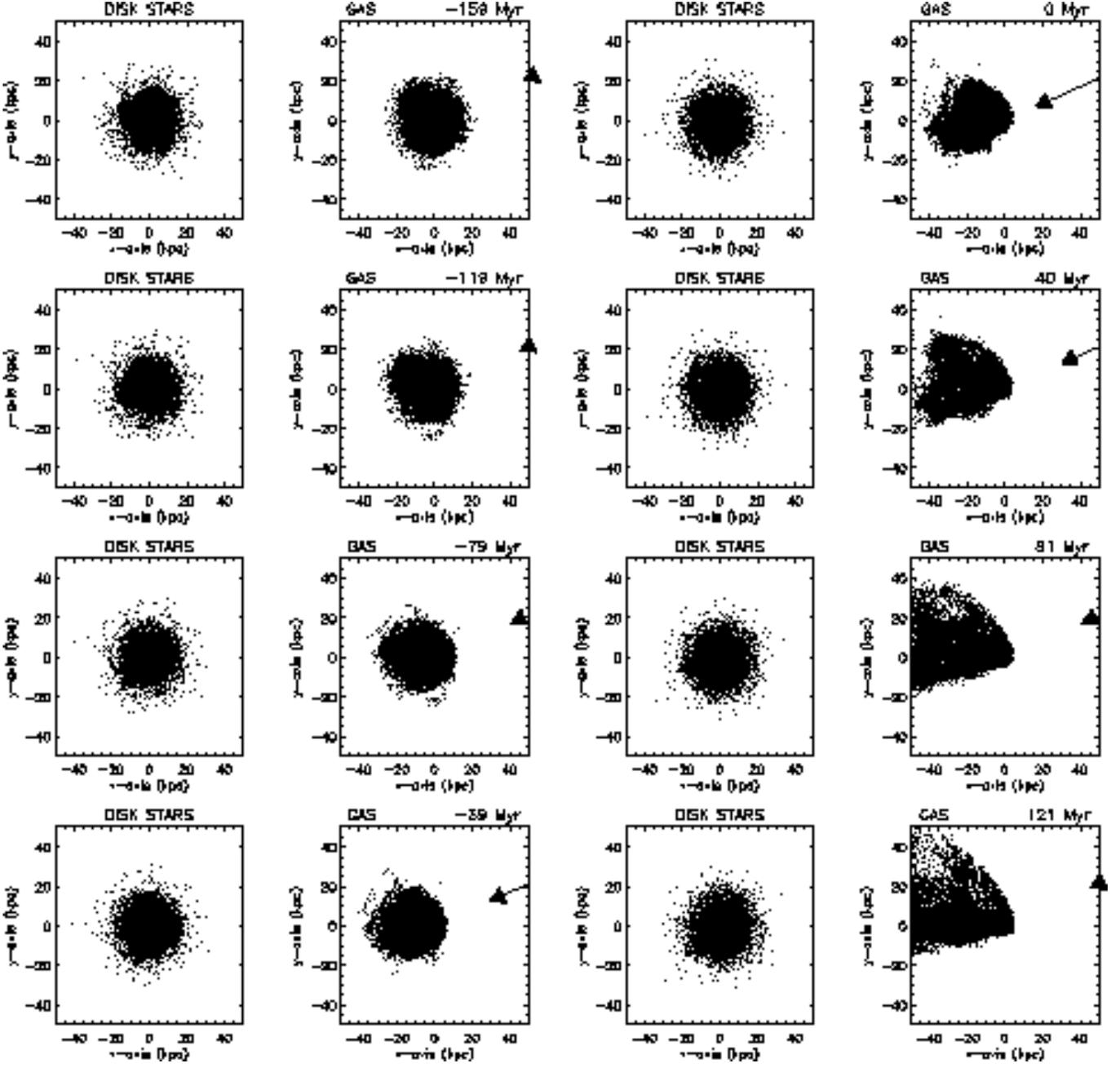}}
        \caption{Evolution of the model stellar and gas disk. 
	The disk is seen face-on and rotates counter-clockwise. The arrow
	indicates the direction of ram pressure, i.e. it is opposite to
	the galaxy's velocity vector. The size of the arrow is proportional
	to $\rho v_{\rm gal}^{2}$. The galaxy passes the cluster core at
	0~Myr. The time to the core passage is marked at the top of the panels.
        } \label{fig:n4388_evolution1}
\end{figure*}

\begin{figure*}
        \resizebox{\hsize}{!}{\includegraphics{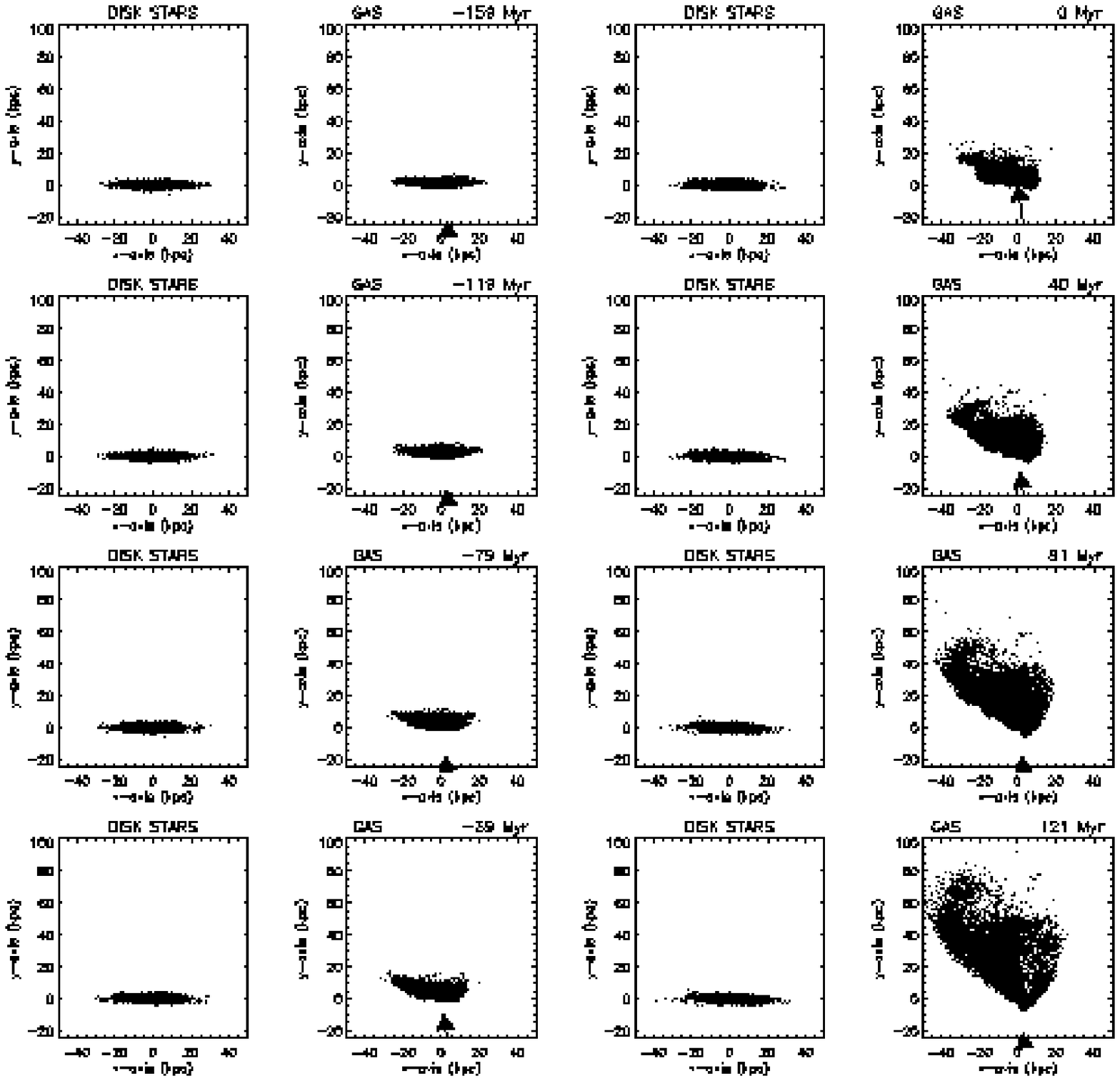}}
        \caption{Evolution of the model stellar and gas disk. 
	The position and inclination angle of NGC~4388 ($PA=90^{\rm o}$, $i=85^{\rm o}$)
	are used. The arrow
	indicates the direction of ram pressure, i.e. it is opposite to
	the galaxy's velocity vector. The size of the arrow is proportional
	to $\rho v_{\rm gal}^{2}$. The galaxy passes the cluster core at
	0~Myr. The time to the core passage is marked at the top of the panels.
        } \label{fig:n4388_evolution}
\end{figure*}

Fig.~\ref{fig:n4388_evolution1} and Fig.~\ref{fig:n4388_evolution} show 
the evolution of the stellar and
gas disk during the galaxy's passage through the cluster core.
The galaxy is seen face-on in Fig.~\ref{fig:n4388_evolution1}.
In  Fig.~\ref{fig:n4388_evolution} the position and inclination
angle of NGC~4388 ($PA=90^{\rm o}$, $i=85^{\rm o}$) are used.
The time of closest approach to the cluster center, i.e. maximum
ram pressure, is $t=0$~Myr. 
Since only the gas is affected by ram pressure, the stellar disk does not
change during the whole simulation. 
At $t \sim -80$~Myr the external $\sim$5~kpc ring of neutral gas begins to be pushed
to the north (Fig.~\ref{fig:n4388_evolution}).
At maximum ram pressure ($t=0$~Myr) this northern extraplanar gas
shows 2 distinct features: (i) an approximately uniform gas layer in the
north of the disk and (ii) a horizontal gas ridge located to the north east.
This is where the accelerated gas, which was initially rotating against the direction of 
the galaxy's velocity (at (0~kpc, 20~kpc) in Fig.~\ref{fig:n4388_evolution1})
and already rotated $\sim$180$^{\rm o}$, is now
again compressed by ram pressure (at (-20~kpc, 0~kpc) in Fig.~\ref{fig:n4388_evolution1}). 
These structures expand in the following
evolution of the neutral gas. For $t>40$~Myr a counter-rotating arm is
forming (see Vollmer et al. 2001), which is responsible for the diagonal 
gas feature in the south-west (Fig.~\ref{fig:n4388_evolution}).

\begin{figure}
        \resizebox{\hsize}{!}{\includegraphics{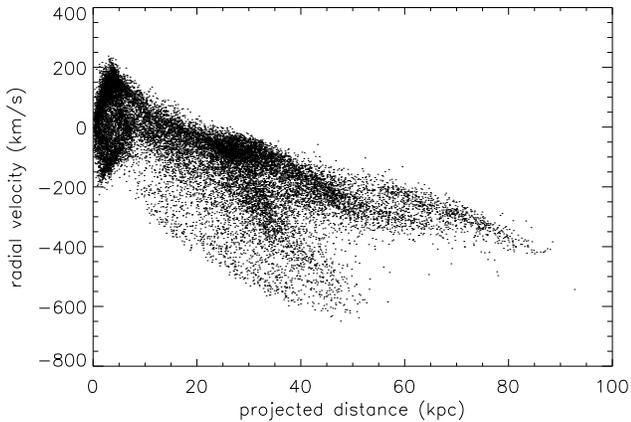}}
	\caption{ISM radial velocity as a function of the projected distance to the
	galaxy center.
	 } \label{fig:r_v_diagram}
\end{figure}

In order to show the effect of ram pressure on the radial velocities,
we show the latter as a function of the projected distance from the galaxy center
for the last snapshot ($t \sim 120$~Myr) in Fig.~\ref{fig:r_v_diagram}.
From 0 to 30~kpc the radial velocities with respect to the galaxy center
show an almost constant gradient from
100~km\,s$^{-1}$ to -50~km\,s$^{-1}$. At a distance of $\sim$30~kpc the
velocities bifurcate. One part continues with the same gradient, the other 
part shows a much steeper gradient. The minimum velocities are -600~km\,s$^{-1}$.
The minimum absolute velocities are $|v| \sim$800~km\,s$^{-1}$.
The main gas mass is found at a projected distances between 20~kpc and 30~kpc
and radial velocities between -50~km\,s$^{-1}$ and 0~km\,s$^{-1}$.
At this timestep, the galaxy has an absolute velocity of $\sim$2000~km\,s$^{-1}$
with respect to the cluster mean velocity. From $p_{\rm max}=5000$~cm$^{-3}$(km/s)$^{2}$
we derive a maximum ICM density of $n_{\rm ICM}^{\rm max} \sim 3\,10^{-3}$~cm$^{-3}$.
Using the ICM density profile of Schindler et al. (1999), this corresponds to an 
impact parameter of $b \sim 100$~kpc, which is consistent with the dynamical
model of Vollmer et al. (2001). We compare the final timestep ($t \sim 120$~Myr)
with our observations.

\section{Discussion \label{sec:discussion}}

Despite the fact that the 100-m spectra do not give precise informations 
about the location of the
detected H{\sc i} gas, we can conclude from our data that the component around
zero radial velocity lies at a distance of at least half a beamsize (4.7$'$ or 23~kpc) 
away from the galaxy center. It contains several 10$^{7}$ solar masses. If it was closer
to the galaxy one should see it in the VLA synthesized spectrum if it is high column density 
gas or alternatively in the central 100-m spectrum if it was low column density, extended gas. 
We suggest that it is connected to the very extended H$\alpha$ plume discovered by Yoshida 
et al. (2002).

To our knowledge this is the first time that atomic gas is observed at these large distances
from a galaxy in a cluster core region. What is the origin of this extraplanar gas?  

Veilleux et al. (1999) discussed 4 different mechanisms (cf. Sect.~\ref{sec:intro}) and favoured
a mixture of nuclear outflow and ram pressure stripping. Yoshida et al. (2002) prefered
a minor merger and/or ram pressure stripping. If the detected atomic gas is connected to
the very extended H$\alpha$ plume, our data rule out a nuclear outflow as
the origin of this plume, because of its large total gas mass. 
A superwind that has recently ceased could have provided enough energy
(see e.g. Heckman et al. 1990),
but there is no indication for a recent, strong  starburst in the center of NGC~4388.
If the atomic gas and the very extended H$\alpha$ plume represent the debris of a
minor merger, its gas velocity around relative zero places its orbit in the plane of the sky.
In our understanding this is inconsistent with the morphology of the H$\alpha$ plume.
If the H$\alpha$ emission represents the debris of a tidally disrupted
dwarf galaxy, it should trace its trajectory.
In this picture the curvature of the north-eastern part of the very extended 
H$\alpha$ emission is dynamically difficult to explain.
In addition, the H$\alpha$ plume is larger close to the disk, which is also
in contradiction to the tidal debris scenario.

\begin{figure*}
        \resizebox{\hsize}{!}{\includegraphics{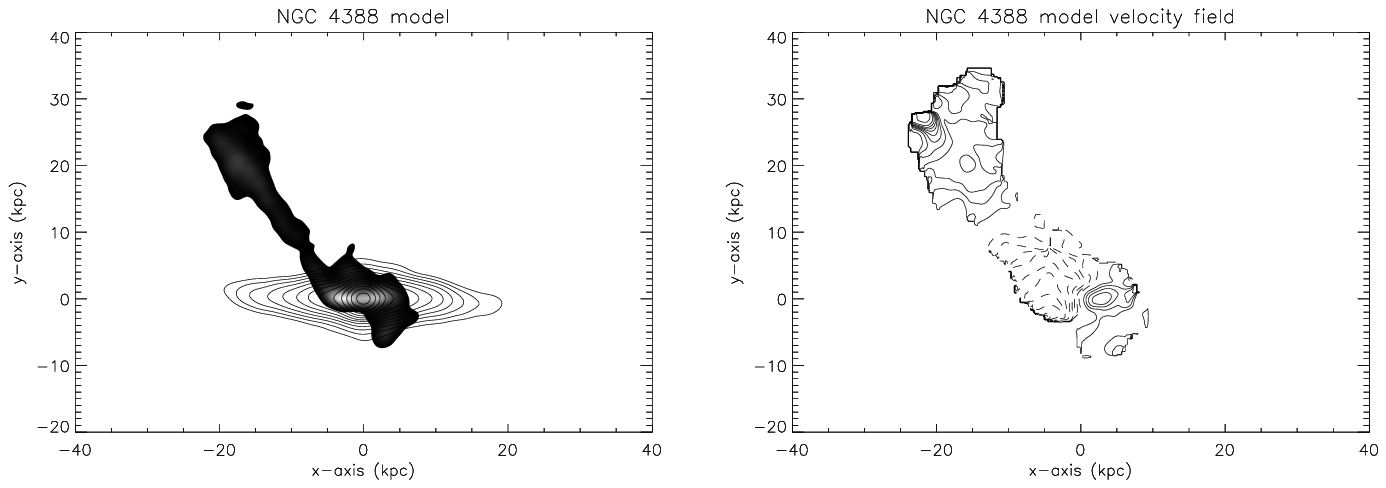}}
        \caption{Left panel: stellar distribution as contours, gas distribution as greyscales.
	Only gas with a local density greater than $\sim 30$~cm$^{-3}$ is shown.
	For clarity the greyscales are inverted, i.e. dark regions correspond to
	low column density gas. Right panel: model velocity field. Dashed lines mark positive,
	solid lines negative velocity contours. The velocities range between
	-200~km\,s$^{-1}$ and 260~km\,s$^{-1}$. The contour stepsize is 20~km\,s$^{-1}$.
        } \label{fig:n4388_model}
\end{figure*}

We suggest a model, where the atomic gas and the H$\alpha$ are connected and both represent
ram pressure stripped gas in different phases. In Fig.~\ref{fig:n4388_model} (left panel)
we show a possible ram pressure stripping scenario. The galaxy moves to the south and away 
from the observer. It is on a radial orbit and its core passage is $\sim$120~Myr ago
(Sect.~\ref{sec:model}). The galaxy has lost $\sim$90\% of its atomic gas. 
We show only gas with a local density greater than $n \sim 30$~cm$^{-3}$.
For clarity the greyscales are inverted, i.e. dark regions correspond to
low column density gas. Since the stars are not affected by ram pressure,
the stellar disk is symmetric. The gas disk is strongly truncated at a radius
of $\sim$5~kpc ($\sim 1'$). This is in agreement with the VLA observations of
Cayatte et al. (1990). Only the high density eastern edge of the stripped, extraplanar
gas distribution can be seen. In addition, the counter-rotating, backfalling arm
in the south-west has high local densities and thus appears in Fig.~\ref{fig:n4388_model} 
(left panel).  The main off-plane gas mass is located at (-20~kpc, 20~kpc).
For simulations with smaller maximum ram pressure, it is located closer to the galaxy.
The corresponding velocity field is shown in Fig.~\ref{fig:n4388_model} (right panel).
Within the disk plane, the velocity field is dominated by rotation.
The western side of the disk is approaching the observer. The stripped gas
at distances smaller than 10~kpc above the disk plane has positive velocities,
whereas the gas at distances greater than 10~kpc has negative velocities.
The counter-rotating, backfalling arm has a mean radial velocity of $\sim 0$~km\,s$^{-1}$.

\begin{figure*}
        \resizebox{\hsize}{!}{\includegraphics{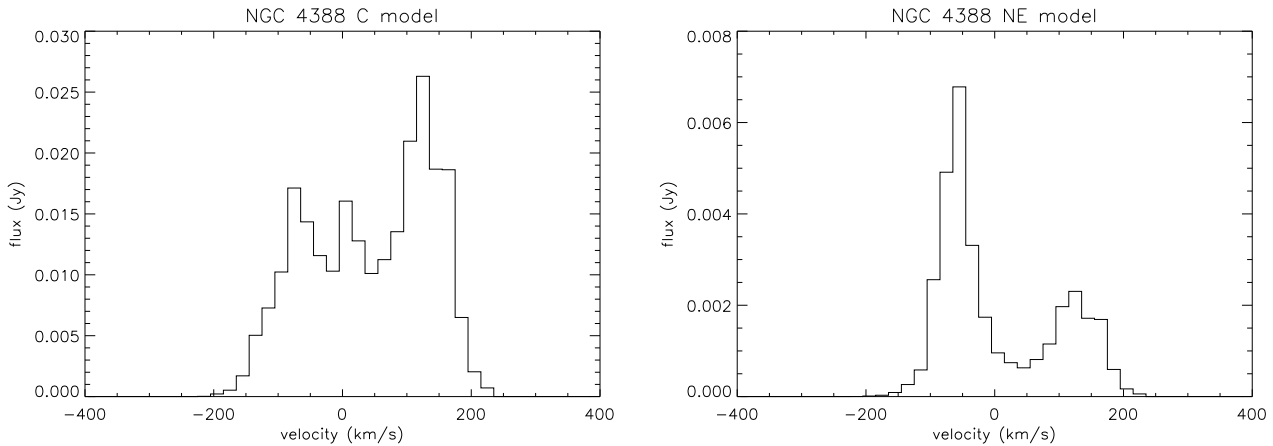}}
        \caption{Left panel: model 100-m spectrum at the central (C) position.
	Right panel: model 100-m spectrum at the northeastern (NE) position.
	Velocities are given relative to the systemic velocity of NGC~4388.
        } \label{fig:n4388_model_spectra}
\end{figure*}

Fig.~\ref{fig:n4388_model_spectra} shows the model 100-m spectrum at central position 
(left panel) and the northeastern (NE) position (right panel). 
We find a total H{\sc i} mass of $\sim 5\,10^{8}$~M$_{\odot}$  
and $\sim 7\,10^{7}$~M$_{\odot}$ at the central and north-eastern position, respectively.

The central spectrum shows a double horn structure and the receding side has
a higher peak flux as it is observed (Fig.~\ref{fig:n4388c}). However, a part of
the flux from the high velocity end of the approaching side is missing, mainly because 
it has a lower local density.

For the north-eastern position we observe a
clear resemblance between the model and the observed spectrum (Fig.~\ref{fig:n4388_spectra}).
The maximum at 160~km\,s$^{-1}$ is due to the extended eastern, receding side, whereas the
main maximum around $\sim$-50~km\,s$^{-1}$ radial velocity is due to the material at distances 
of about 20~kpc from the galactic disk (Fig.~\ref{fig:n4388_model}). 
Thus the ram pressure scenario can account for the
very extended tail. However, it is not possible to reproduce the velocity field
of the NE H$\alpha$ plume (Veilleux et al. 1999), since it shows
a blueshift with respect to the galaxy's systemic velocity. 
In order to explain this feature an additional outflow mechanism is needed.

How can this gas survive within such a hostile environment like the intracluster medium (ICM)?
Our model is not able to investigate this subject, because it considers only
the neutral gas phase and does not include ISM/ICM shocks that might heat the gas.
Evaporation and ionization physics are also not included. Thus we have to rely on
analytical arguments.

Veilleux et al. (1999) argued 
that in the case classical thermal heat conduction a typical H{\sc i} cloud evaporates in 
the hot ICM within 10$^{7}$~yr. On the other hand, Vollmer et al. (2001) showed 
that within the ICM evaporation is saturated. If the stripped atomic gas clouds have column 
densities of 10$^{21}$~cm$^{-2}$, the evaporation time can reach $\sim$10$^{8}$~yr. They also
discussed the possibility that the dense part of the stripped neutral gas contracts
due to its selfgravity, because the internal heating (stellar radiation and supernova shocks)
are absent. Since the dust is stripped together with the gas, molecules might form
when the gas density becomes high enough. The only gas heating is then the X-ray radiation
from the hot ICM. An equilibrium cloud has a mass of several thousand solar masses, a radius
of $\sim$10~pc, and a column density of a few $10^{21}$~cm$^{-2}$. Only 10\% of this
column density is in form of neutral hydrogen. Could this be the case for NGC~4388?
These clouds can in principle survive long enough to be pushed to the observed large distances.
It is then not excluded that these clouds collapse and form stars.

Interestingly, Gerhard et al. (2002) found a compact H{\sc ii} region that has formed 
stars in situ, far away from the main star-forming disk. This H{\sc ii} region is located
at a projected distance of 17.5~Mpc from the galaxy center
in the northwest of the galactic center and has a radial velocity of 150~km\,s$^{-1}$.
Within our model it is possible that this H{\sc ii} region is falling back to the galaxy after 
having been pushed by ram pressure to a large galactocentric distance. 
When a stripped cloud collapses, its column density increases rapidly and
it decouples from the wind (ram pressure). It then begins to fall back to the galaxy
due to the galaxy's gravitational potential. 
With the relatively low metallicity (Gerhard et al. 2002) this stripped gas was located 
approximately
at the optical radius ($R_{25}$) of NGC~4388 before stripping has taken place (Duc et al. 2000).
This scenario is in agreement with the positive radial velocity of the compact H{\sc ii} region.
In addition, Gerhard et al. (2002)  found a number of H{\sc ii} region candidates in the
very extended H$\alpha$ plume. Within our picture this would mean that the stripped gas
clouds have become molecular and are now forming stars. This mechanism would be analog to the
formation of globular clusters in the tidal tails of gravitationally perturbed galaxies.

\section{Conclusions \label{sec:conclusions}}

We have observed the Virgo cluster core galaxy NGC~4388 in the 21~cm H{\sc i}
line with the Effelsberg 100-m telescope at 5 different positions: 
one in the center and 4 located one 100-m
beamsize to the NE, NW, SW, and SE from the galaxy center. The results of these
observations are compared to a realistic dynamical model including ram pressure stripping.
The results are:
\begin{enumerate}
\item
We detected extraplanar H{\sc i} gas of $\sim$6\,10$^{7}$~M$_{\odot}$ located at a distance 
of at least 20~kpc to the north-east of the galaxy center. This gas is most 
probably connected to the very extended H$\alpha$ plume detected by Yoshida et al. (2002).
\item
The dynamical model is able to reproduce the H{\sc i} deficiency, the truncated H{\sc i} disk 
and the extraplanar emission. NGC~4388 passed the cluster core $\sim$120~Myr ago.
\item
Within this scenario the maximum ram pressure is $p_{\rm max}=5000$~cm$^{-3}$(km/s)$^{-2}$,
the maximum ICM density encountered during core passage is $n_{\rm ICM} \sim 3\,10^{-3}$~cm$^{-3}$,
the galaxy's absolute velocity with respect to the cluster mean is $\sim$2000~km\,s$^{-1}$,
and the angle between the disk and the orbital plane is $i = 45^{\rm o}$
\item
The extraplanar compact H{\sc ii} region recently found by Gerhard et al. (2002) can be 
explained as a stripped gas cloud that collapsed and decoupled from the ram pressure wind
due to its increased surface density.
\end{enumerate}

\begin{acknowledgements}
Based on observations with the 100-m telescope of the MPIfR (Max-Planck-Institut f\"{u}r 
Radioastronomie) at Effelsberg.
The authors would like to thank V. Cayatte and J. van Gorkom for providing us
kindly their VLA data.
\end{acknowledgements}

\end{document}